\documentclass[prd,preprint,showpacs,eqsecnum,floatfix]{revtex4}

\usepackage{amsfonts}
\usepackage{amssymb}
\usepackage{graphicx}
\usepackage{pstricks}
\usepackage{multirow}

\newcommand{\be}{\begin{equation}}
\newcommand{\ee}{\end{equation}}
\newcommand{\beq}{\begin{eqnarray}}
\newcommand{\eeq}{\end{eqnarray}}
\newcommand{\bpi}{\begin{picture}}
\newcommand{\bce}{\begin{center}}
\newcommand{\epi}{\end{picture}}
\newcommand{\ece}{\end{center}}

\def\lsim{\mathrel{\rlap{\lower4pt\hbox{$\sim$}}
    \raise1pt\hbox{$<$}}}                
\def\gsim{\mathrel{\rlap{\lower4pt\hbox{$\sim$}}
    \raise1pt\hbox{$>$}}}                

\begin{document}

\begin{flushright}
FTUV-0603-30\\[-2mm]
ECT*-06-03\\[-2mm] 
\end{flushright}

\title{CP violation through particle mixing and the 
$H$-$A$ lineshape}

\author{J.~Bernab\'eu$^a$, D.~Binosi$^{b}$ and J.~Papavassiliou$^a$ }

\affiliation{
$^a$ECT*, Villa Tambosi, Strada delle Tabarelle 286
I-38050 Villazzano (Trento), Italy\\
$^b$Departamento de F\`\i sica Te\`orica, and IFIC
Centro Mixto, Universidad de Valencia-CSIC, E-46100, Burjassot,
Valencia, Spain}

\begin{abstract}

We  consider  the possibility  of  looking  for  CP-mixing effects  in
two-Higgs doublet  models (and particularly  in the MSSM)  by studying
the lineshape of  the CP-even ($H$) and CP-odd  ($A$) neutral scalars.
In most  cases $H$ and  $A$ come quite  degenerate in mass,  and their
$s$-channel  production would lead  to nearly  overlapping resonances.
CP-violating effects may connect these two Higgs bosons, giving origin
to one-loop particle  mixing, which, due to their  mass proximity, can
be  resonantly  enhanced.    The  corresponding  transition  amplitude
contains  then CP-even and  CP-odd components;  besides the  signal of
intereference between both amplitudes,  leading to a CP-odd asymmetry,
we  propose to  look for  the  mixing probability  itself, a  quantity
which, although  CP-even, can originate only from  a CP-odd amplitude.
We  show that, in  general, the  effect of  such a  mixing probability
cannot be mimicked  by (or be re-absorbed into)  a simple redefinition
of the  $H$ and $A$  masses in the  context of a  CP-conserving model.
Specifically, the effects  of the CP-mixing are such  that, either the
mass-splitting  of the  $H$  and  $A$ bosons  cannot be accounted for 
in the absence of CP-mixing, and/or the detailed
energy dependence of the  produced lineshape is clearly different from
the  one obtained  by  redefining  the masses,  but  not allowing  any
mixing.   This  analysis  suggests  that  the detailed  study  of  the
lineshape of this Higgs system may provide valuable information on the
CP nature of the underlying theory.

\end{abstract}


\maketitle

\section{Introduction}

Despite  the  impressive  success   of  the  Standard  Model  (SM)  in
describing a  plethora of high-precision experimental  data \cite{PDB}, 
a number
of  important  theoretical  issues  related to  its  structure  remain
largely  unexplored. Perhaps  the  most prominent  among  them is  the
nature  of the very  mechanism which  endows the  elementary particles
with their observed masses.   The best established way for introducing
masses  at tree-level, without  compromising renormalizability,  is 
the Higgs mechanism \cite{Higgs:1964ia,Englert:1964et,Guralnik:1964eu}, 
where  the crucial ingredient  is the
coupling of all  would-be massive particles to a  complex scalar field
with   a   non-vanishing   vacuum   expectation   value.    The   most
characteristic physical remnant of  this procedure is a massive scalar
particle in the spectrum of  the theory, the Higgs boson.  In addition
to the  minimal SM,  the Higgs mechanism  is employed in many 
popular  new  physics scenaria,  most  notably  in  the 
Minimal Supersymmetric Standard Model (MSSM) and  its
variants \cite{Nilles:1983ge}, 
leading  to yet richer scalar sectors.  However,  to date,
neither the Higgs boson of  the minimal SM, nor the additional scalars
predicted  by its  supersymmetric extensions  have been  observed, and
their discovery and subsequent detailed study 
is considered as one  of the main  priorities for the upcoming 
collider experiments.

Several of the aforementioned scalars, and especially the 
``standard'' Higgs boson, 
 are expected to be discovered at the LHC~\cite{LHC}. In addition,
in the past few years  a significant amount of technical and
theoretical research  has  been  invested  in the   evaluation  of the
feasibility and physics  potential of  muon colliders 
\cite{Cline:1994tk,Marciano:1998ue}.  Such  machines
will have   the particularly appealing  feature  of variable centre-of-mass 
energy, which will  allow the resonant enhancement of $s$-channel
interactions in order to produce  Higgs bosons copiously.  Thus,  muon
colliders are  expected to operate as  Higgs  factories, in the energy
range of up to .5 TeV. Given the additional  features of small energy
spread and precise energy determination, these machines offer a unique
possibility  for studying the  line-shape of  the  SM Higgs  boson, and
determining its  mass and width with an  accuracy of about .1 MeV and
.5  MeV, respectively.   In addition, if supersymmetric  particles
will be discovered at the LHC,  then the muon  colliders would play the
role  of a  precision machine  for   studying their main properties,
such as masses, widths, and couplings, and delineating their line-shapes
\cite{Barger:1996jm,Binoth:1997ev,Krawczyk:1997be,Grzadkowski:2000fg,Blochinger:2002hj}.

In the two-Higgs doublet models 
in general~\cite{Lee:1973iz,Branco:1985aq,Liu:1987ng,Weinberg:1990me,Wu:1994ja}, 
and in most SUSY scenarios in 
particular~\cite{Nilles:1983ge}, the extended scalar sector contains five physical fields: a couple of charged Higgs bosons ($H^\pm$), a CP-odd scalar $A$, and two CP-even scalars $h$ (the lightest, which is to be identified with the SM Higgs) and $H$ (the heaviest).
The detailed study of the  
$H$-$A$  system  is  particularly  interesting 
because in most {\it beyond the SM} scenarios (such as SUSY) the masses of these two particles are
almost degenerate, and therefore their resonant  production is expected to
give rise to nearly overlapping resonances~\cite{Barger:1996jm}.

Specifically, the tree-level mass eigenvalues of the Higgs mass matrices are given by
\begin{eqnarray}
m^2_{H^\pm}&=&m^2_A+M^2_W, \nonumber \\
m^2_{h,H}&=&\frac12\left[M^2_Z+m^2_A\mp\sqrt{(M^2_Z+m^2_A)^2-4m^2_AM^2_Z\cos^22\beta}\right],
\end{eqnarray}
and therefore one will have the tree-level bounds $M_Z\in(m_h,M_H)$ and $M_W<m_{H^\pm}$. 
Furthermore, in the decoupling limit $M_A\gg M_Z$, the neutral sector (tree-level) masses satisfy the relations 
~\cite{declim}
\begin{eqnarray}
m^2_h&\approx&M^2_Z\cos^22\beta,\nonumber \\
m^2_H&\approx& m^2_A + M_Z^2 \sin^22\beta,
\label{deg}
\end{eqnarray}
which,  for $\tan\beta\ge2$  (and thus  $\cos^22\beta\approx1$), imply
the degeneracies  $m_h\approx M_Z$ and $m_H\approx  m_A$. 

In general, the inclussion of radiative corrections is known to modify 
significantly 
the above tree-level relations, mainly due to the 
large Yukawa coupling of the top-quark.
Such corrections  are particularly important for the lower bound 
on the mass 
of the lightest Higgs boson $h$:
for the largest  bulk of  the parameter
space, 
$h$ is in fact heavier than the $Z$ boson, and can be as heavy as 130 GeV~\cite{Mh}. 
It is important to emphasize, however, that  
the inclusion of radiative corrections does {\it not} lift 
the mass degeneracy  in the  $H$-$A$ system under consideration, 
especially in the parameter space
region where $m_A>2M_Z$ and  $\tan\beta\ge2$~ \cite{KKRW}.
In particular, as we will see in detail below,
quantum effects do shift the values of $m_H$ and $m_A$, 
but by amounts that are numerically comparable;
as a result, the small mass splitting between $H$ and $A$ 
remains still valid
(but is not described quantitatively by Eq.~(\ref{deg})).

In  addition, as was  shown by  Pilaftsis~\cite{Pilaftsis:1997dr}, the
near  degeneracy of the  $H$-$A$ system  may give  rise to  a resonant
enhancement of  CP violation due to particle  mixing. Specifically, if
the CP  symmetry is exact, the $H$  cannot mix with $A$,  at any given
order.  However,  in  the  presence  of  a  CP-violating  interactions
\cite{Bernabeu:1987gr,Pilaftsis:1992st,Korner:1992zk,Bernabeu:1993up,
Ilakovac:1993pt,Bernabeu:1995ph,Tommasini:1995ii,Pilaftsis:1998dd}, in
addition        to        a        variety       of        interesting
implications~\cite{Pilaftsis:1998pe},  the $H$  can mix  with  the $A$
already  at one-loop  level,  giving rise  to  a non-vanishing  mixing
self-energy  $\Pi_{HA}(s)$.  Such mixing,  in  turn,  can be  measured
through  the   study  of  appropriate  CP-odd   observables,  such  as
left-right     asymmetries.    As     has     been    explained     in
\cite{Pilaftsis:1997dr},  in  general  the CP-violating  amplitude  is
particularly enhanced  near resonace, if the two  mixing particles are
nearly degenerate,  a condition which  is naturally fullfilled  in the
$H$-$A$ system.   Furthermore, the mixing  between $H$ and $A$  has an
additional profound effect for the  two masses: the near degeneracy of
the two  particles is lifted, and  the pole masses  move further apart
\cite{Carena:2001fw};   as  a   result,  the   originally  overlapping
resonances of the CP-invariant theory tend to be separated.

\begin{figure}[t]
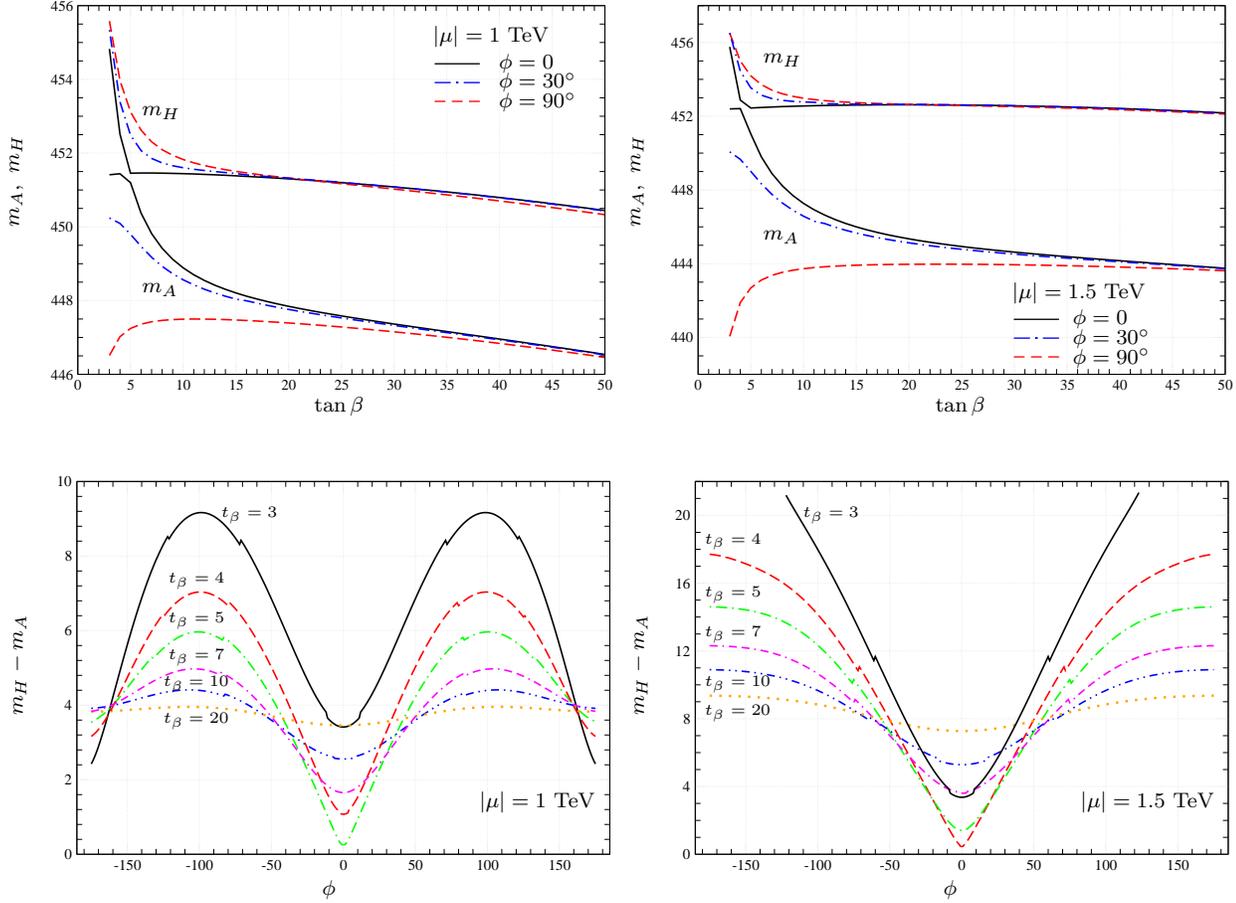


\begin{picture}(300,340)(90,-170)
\begin{tabular}{cc}
\includegraphics[width=7.5cm]{m_vs_tb-1-1.eps} \hspace{10pt} \vspace{25pt} & \includegraphics[width=7.5cm]{m_vs_tb-1.5-1.eps}\\
\includegraphics[width=7.4cm]{dm_vs_phi-1-1.eps} \hspace{12pt} & \includegraphics[width=7.4cm]{dm_vs_phi-1.5-1.eps} \\ 
\end{tabular}
\put(-305,155){\scriptsize{$|\mu|=1$ TeV}}
\psline[linewidth=.5pt](-10.7,5.15)(-10.1,5.15)
\put(-280,145){\scriptsize{$\phi=0$}}
\psline[linewidth=.5pt,linecolor=blue,linestyle=dashed,dash=5pt 4pt](-10.7,4.9)(-10.1,4.9)
\psline[linewidth=.5pt,linecolor=blue,linestyle=dotted](-10.41,4.9)(-10.39,4.9)
\put(-280,137.5){\scriptsize{$\phi=30^\circ$}}
\psline[linewidth=.5pt,linecolor=red,linestyle=dashed](-10.7,4.65)(-10.1,4.65)
\put(-280,130){\scriptsize{$\phi=90^\circ$}}
\put(-415,127){\scriptsize{$m_H$}}
\put(-415,60){\scriptsize{$m_A$}}
\put(-465,85){\rput[lt]{90}{\scriptsize{$m_A,\ m_H$}}}
\put(-350,15){\scriptsize{$\tan\beta$}}
\put(-86,58){\scriptsize{$|\mu|=1.5$ TeV}}
\psline[linewidth=.5pt](-3.0,1.75)(-2.4,1.75)
\put(-63,48){\scriptsize{$\phi=0$}}
\psline[linewidth=.5pt,linecolor=blue,linestyle=dashed,dash=5pt 4pt](-3.0,1.5)(-2.4,1.5)
\psline[linewidth=.5pt,linecolor=blue,linestyle=dotted](-2.71,1.5)(-2.69,1.5)
\put(-63,40.5){\scriptsize{$\phi=30^\circ$}}
\psline[linewidth=.5pt,linecolor=red,linestyle=dashed](-3.0,1.25)(-2.4,1.25)
\put(-63,33){\scriptsize{$\phi=90^\circ$}}
\put(-180,147){\scriptsize{$m_H$}}
\put(-180,80){\scriptsize{$m_A$}}
\put(-230,85){\rput[lt]{90}{\scriptsize{$m_A,\ m_H$}}}
\put(-115,15){\scriptsize{$\tan\beta$}}
\put(-287.5,-135){\scriptsize{$|\mu|=1$ TeV}}
\put(-385,-25){\tiny{$t_\beta=3$}}
\put(-405,-50){\tiny{$t_\beta=4$}}
\put(-405,-65){\tiny{$t_\beta=5$}}
\put(-405,-79){\tiny{$t_\beta=7$}}
\put(-407,-89){\tiny{$t_\beta=10$}}
\put(-407,-103){\tiny{$t_\beta=20$}}
\put(-465,-100){\rput[lt]{90}{\scriptsize{$m_H-m_A$}}}
\put(-347,-168){\scriptsize{$\phi$}}
\put(-60,-135){\scriptsize{$|\mu|=1.5$ TeV}}
\put(-165,-25){\tiny{$t_\beta=3$}}
\put(-202,-35){\tiny{$t_\beta=4$}}
\put(-202,-55){\tiny{$t_\beta=5$}}
\put(-202,-70){\tiny{$t_\beta=7$}}
\put(-202,-89){\tiny{$t_\beta=10$}}
\put(-202,-100){\tiny{$t_\beta=20$}}
\put(-230,-100){\rput[lt]{90}{\scriptsize{$m_H-m_A$}}}
\put(-112,-168){\scriptsize{$\phi$}}
\end{picture}
\caption{\it The mass spectrum of the $H-A$ system in the MSSM with and without CP-breaking effects. Upper panels: the variation of $m_A$ and $m_H$ as a function of $\tan\beta$, for different values of $\phi$ and $|\mu|$; Lower panels: the variation of $\delta m=m_H-m_A$ as a function of $\phi$, for different values of $\tan\beta$ and $|\mu|$ (spikes appearing in the lower panels are code artifacts and should be ignored).}
\label{spectrum}

\end{figure}

Using the  code \verb|CPsuperH|  \cite{Lee:2003nt}, one can  study the
$H-A$  system  mass  spectrum   in  the  CP-invariant  ($\phi=0$)  and
CP-breaking ($\phi\ne0$) limits of the  MSSM. For our purposes we have
assumed  (motivated by  SUGRA  models) universal  squarks soft  masses
$M_0$  (with  $M_0=.5$  TeV),  trilinear couplings  $A$  (with  $\vert
A\vert=1$  TeV) and CP  breaking phases  $\phi_A=\phi$, with  the MSSM
$\mu$ parameter real  and varying between 1.0 and  1.5 TeV. Finally we
set $m_{H^\pm}=.4571$ TeV which sets the tree-level mass of the CP-odd
Higgs  at the value  of .450  TeV.  In  the first  two panels  of Fig.
\ref{spectrum}, we  plot the neutral CP-even and  CP-odd scalar masses
as a function of $\tan\beta$ for three values of the CP-breaking phase
and two different  values of $\mu$ ($\vert\mu\vert=1$ TeV  in the left
panel,  while $\vert\mu\vert=1.5$  TeV in  the right  one).   When the
CP-breaking  phases are  absent (black  continuous curves),  the $H$-$A$
resonance  can  be clearly  seen  at  $\tan\beta=5$ and  $\tan\beta=4$
respectively.  The  degeneracy is lifted when the  phases are switched
on  (blue  dashed-dotted  and  red  dahsed  curves,  corresponding  to
$\phi=30^\circ$  and  $\phi=90^\circ$  degrees,  respectively),  their
effect  becoming rapidly  smaller for  larger $\tan\beta$  values: from
$\tan\beta\gsim 20$ they cannot be disentangled from radiative effects
calculated in  the CP-invariant limit of  the theory. A  more in depth
view of  how the CP-breaking  phases affect the $H$-$A$  mass spectrum
can be  achieved by studying  the mass difference $\delta  m= m_H-m_A$
versus $\phi$ for  different values of $\tan\beta$, as  it is shown in
the   two    lower   panels   of    Fig.\ref{spectrum}   (as   before,
$\vert\mu\vert=1$ TeV in the left panel, while $\vert\mu\vert=1.5$ TeV
in the right  one). In both cases we  see that when $\tan\beta\gsim20$
(orange  dotted lines)  the  difference between  the CP-invariant  and
CP-breaking  case  is barely  visibile.  There  are some  differences,
however, depending on the value  of the $\mu$ parameter. For $|\mu|=1$
TeV,  the mass  difference  shows a  maximum around  $\phi=100^\circ$;
moreover in  the CP-invariant limit  the mass splitting is  bigger for
lower values of $\tan\beta$. When $|\mu|=1.5$ TeV, instead, $\delta m$
has no maximum  and in the CP-invariant limit  the splitting is bigger
for larger $\tan\beta$ values.

The above analysis  suggests that, when studying the  lineshape of the
$H$-$A$  system,  one may  envisage  two,  physically very  different,
scenarios.  In the  first  one, the  CP  symmetry is  exact, with  the
position of resonances determined by Eq.(\ref{deg}) plus its radiative
corrections; the relative position between the two resonance will then
specify $\tan\beta$. In the second scenario, CP is violated, resulting
in mixing  at one-loop  level between CP-even  and CP-odd  states wich
translates  into   a  non-vanishing  off-shell   transition  amplitude
$\Pi_{HA}(s)$. Then, for the same  mass splitting $\delta m$ as in the
previous case, one may not reach  the same conclusion for the value of
$\tan\beta$, because  one could have  started out with the  two masses
almost degenerate, corresponding to  a different value for $\tan\beta$
(lower or higher  depending on the value of  the $\mu$ parameter), and
the observed separation between $m_{H}$  and $m_{A}$ may be due to the
lifting   of  the  degeneracy   produced  by   the  presence   of  the
aforementioned $\Pi_{HA}(s)$.

To address  these questions in  detail, we then  turn to the  MSSM, in
which the one-loop mixing between the CP-even and CP-odd Higgs will be
due   to   third-family   squarks   circulating   in   the   loop   of
Fig.\ref{feyndiag},  and study  the $H$-$A$  lineshape, which  will be
derived   explicitly  at   one-loop  level   for  general   values  of
$\Pi_{HH}(s)$, $\Pi_{AA}(s)$, and  $\Pi_{HA}(s)$. Our main results may
be then summarized as follows.  In the cases for which the CP-breaking
phases are sizeable  (we will set $\phi=90^\circ$ for  this case), one
can always clearly distinguish between the CP-invariant and CP-beaking
scenarios, either  because the  mass splitting in  the latter  case is
just too big for being due to CP-invariant radiative corrections only,
or  because   of  the  different   energy  dependence  of   the  cross
section.   For  smaller   values  of   the  CP-breaking   phases  (say
$\phi=30^\circ$), one  can still distinguish  between the CP-invariant
and   CP-breaking  case   only   when  $\tan\beta$   is  small;   when
$\tan\beta\gsim10$  one cannot tell the two cases apart simply by 
studying the lineshape.

We therefore conclude that the experimental determination of the Higgs
line-shape, in conjunction with  CP-odd asymmetries and other suitable
observables, may  provide valuable information for  settling the issue
regarding CP-mixing effects in two-Higgs models.

\section{The line-shape in the presence of H-A mixing}

In this section we will concentrate on the calculation of the line-shape for the
resonant process $\mu^+\mu^-\to A^*,\ H^*\to f\bar f$ in the presence of the 
CP-violating one-loop mixing diagrams of Fig.1 (a) between the CP-even ($H$) 
and the CP-odd ($A$) Higgses.
\begin{figure}
\includegraphics[width=16cm]{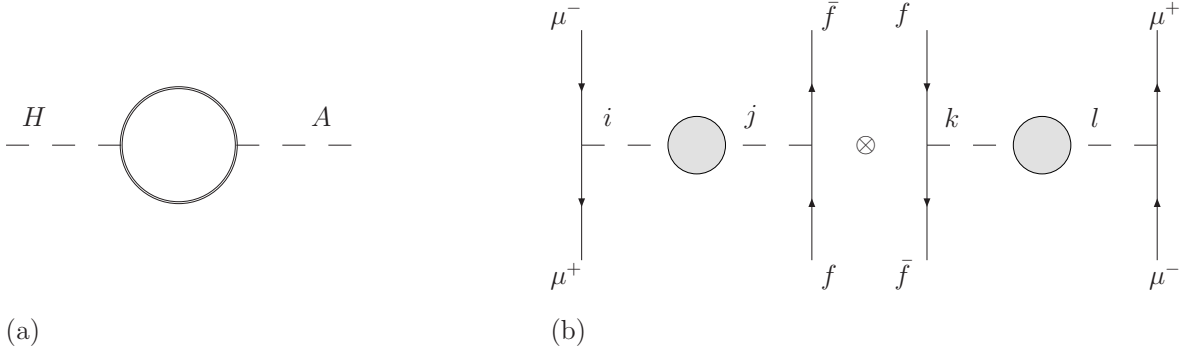}
\caption{(a) {\it The one-loop $AH$ mixing responsible for the CP-violating effects. Particles
circulating in the loop may vary according to the model (in our case they are either
squarks or heavy Majorana neutrinos)}. (b) \it The cross section for $\sigma^{ij}_{kl}$ the resonant 
process $\mu^+\mu^-\to A^*,\ H^*\to f\bar f$. In the resonant region and at the
one-loop level the only diagram contributing is the one shown. Notice that ${i,j,k,l}$ 
may be either $A$ or $H$ (with
$A$-$H$ mixing due to the one-loop diagram (a)); however due to the chirality difference
between  $A$ and $H$, the only non-zero combinations are obtained when $i=l$ and $j=k$.} 
\label{feyndiag}
\end{figure}
The amplitude for such process can be written as
\begin{equation}
T=T^s+T^t=V_i^P\widehat\Delta_{ij}V_j^D+T^t,
\end{equation}
where $i,j=A,H$, $V_i^P$ ($V_j^D$) is the production (decay) amplitude of the process $\mu^+\mu^-\to i$ 
($j\to\bar f f$), $\widehat\Delta_{ij}$ is the propagator matrix in the presence of mixing, 
and $T^t$ represents collectively all the $t$-channel and box diagrams. 

The propagator matrix $\widehat\Delta_{ij}(s)$ appearing in the equation above is constructed from self-energies
resummed in the pinch technique framework \cite{Papavassiliou:1995fq}; 
they are gauge-independent, 
display only physical thresholds, have the correct unitarity, analiticity, and
renormalization group properties, and satisfy the equivalence theorem. In particular, one has
\begin{eqnarray}
\widehat\Delta_{AA}(s) &=& [s-M_H^{2} + \widehat\Pi_{HH} (s)] \widehat\Delta^{-1}(s), \nonumber\\
\widehat\Delta_{HH}(s) &=& [s-M_A^{2} + \widehat\Pi_{AA} (s)] \widehat\Delta^{-1}(s), \nonumber\\
\widehat\Delta_{HA}(s) &=& \widehat\Delta_{AH}(s) = - \widehat\Pi_{AH}(s) \widehat\Delta^{-1}(s),
\end{eqnarray}
with $\widehat\Delta(s)$ the determinat of the propagator matrix, i.e.,
\begin{equation}
\widehat\Delta(s)=[s-M_H^{2} + \widehat\Pi_{HH} (s)][s-M_A^{2} + \widehat\Pi_{AA} (s)] - \widehat\Pi_{AH}(s) \widehat\Pi_{HA}(s).
\end{equation}
In the resonant region on which we concentrate here, $M_A\approx M_H$, the $t$-channel and box
diagrams are subdominant, with $T^t\ll T^s$, and can be safely ignored; moreover, at the one-loop level, 
the production and decay amplitudes $V^P_i$ and $V^D_j$ coincide with the corresponding tree-level (model dependent) vertices    
$\Gamma^{(0)}_{\mu^+\mu^-i}=\chi_{\mu i}$ and $\Gamma^{(0)}_{jf\bar f}=\chi_{jf}$. Finally, in the resonant region, 
the dispersive (real) part of the self-energies are very small, so that one can neglect them and concentrate only on 
their absorptive (imaginary) parts. 
The cross sections leading to the lineshape we look for, can then be calculated by just squaring the single diagram shown
in Fig. \ref{feyndiag} (b); helicity mismatches limit the possible combinations to the factors
\begin{eqnarray}
D_{AA}^{AA}(s) &=& \left[(s-M_H^{2})^{2} + (\mathrm{Im}\;\!\widehat\Pi_{HH} (s))^{2}\right] 
D^{-1}(s), \nonumber\\
D_{HH}^{HH}(s) &=& \left[(s-M_A^{2})^{2} + (\mathrm{Im}\;\!\widehat\Pi_{AA} (s))^{2}\right]
D^{-1}(s)\, , \nonumber\\
D_{HA}^{AH}(s) &=& D_{AH}^{HA}= \widehat\Pi_{AH}^{2}(s) D^{-1}(s),
\end{eqnarray}
with
\begin{eqnarray}
D(s) &=& \left[(s-M_A^{2})(s-M_H^{2}) - \mathrm{Im}\;\!\widehat\Pi_{AA} (s) 
\mathrm{Im}\;\!\widehat\Pi_{HH} (s) - \widehat\Pi_{AH}^{2}(s)\right]^{2} \nonumber\\ 
&+&  \left[(s-M_A^{2})\mathrm{Im}\;\!\widehat\Pi_{HH}(s) + (s-M_H^{2})\mathrm{Im}\;\!\widehat\Pi_{AA}(s)\right]^{2}.
\end{eqnarray}
In the formulas above, the standard thresholds are given by
\begin{eqnarray}
\mathrm{Im}\;\!\widehat\Pi_{HH}^{(f\bar{f})}(s) &=& \frac{\alpha_\mathrm{w} N^f_c}{8}
\chi_{Hf}^2 \frac{m^2_f}{M^2_W} s\
\left(1-\frac{4m^2_f}{s}\right)^{3/2} \theta (s-4m^2_f ),\nonumber \\
\mathrm{Im}\;\!\widehat\Pi_{AA}^{(f\bar{f})}(s) &=& \frac{\alpha_\mathrm{w} N^f_c}{8}
\chi_{Af}^2 \, \frac{m^2_f}{M^2_W}\, s\,
\left(1-\frac{4m^2_f}{s}\right)^{1/2} \theta (s-4m^2_f ),\nonumber \\
\mathrm{Im}\;\!\widehat\Pi_{HH}^{(VV)}(s) &=&
\frac{n_V\alpha_\mathrm{w}}{32} \chi_{HV}^2 \frac{M^4_H}{M^2_W}
\left(1-\frac{4M^2_V}{s}\right)^{1/2}\nonumber\\ 
&\times& \left[
1+4\frac{M^2_V}{M^2_H}-4\frac{M^2_V}{M^4_H}(2s-3M^2_V) \right] \theta
(s-4M^2_V ),
\label{thresh}
\end{eqnarray}
where, $\alpha_\mathrm{w}=g^2/4\pi$, $n_V=2$ (respectively 1) for $V\equiv W$ 
(respectively $V\equiv Z$), and $N^f_c=1$ for leptons and 3 for quarks.
Restoring the couplings, summing over final states and averaging over initial polarizations, 
one is then left with the cross-sections
\begin{equation}
\sigma(s)=\sum_{i,j}\sigma^{ij}_{ji}(s)
=  \frac{\pi \alpha_\mathrm{w}^2 m_{\mu}^2 m_{b}^2}{16 M_W^4}s 
\sum_{i,j}\chi_{i\mu}^{2} \chi_{jb}^{2} D_{ji}^{ij}(s),
\label{cs}
\end{equation}
where we have specialized to the case in which the final fermions are bottom quarks.

\section{Numerical results}

In this section we present numerical results for the cross section of Eq.(\ref{cs}), in order 
to see how CP violating effects through the one-loop mixing of Fig.\ref{feyndiag} (a) affects the Higgs line-shape. 

As anticipated in the introduction, the  MSSM the  Higgs  sector is  composed  by  two Higgs  doublets
$(A_1,A_2)$ and $(H_1,H_2)$.  After performing an orthogonal rotation by the angles $\beta$ (with  $\tan\beta=v_2/v_1$ and $v_i$ the VEVs of
the Higgs doublets)  and $\alpha$ (with $\tan\alpha\propto\tan\beta$),
the doublets end up in  the physical basis $(G^0,A)$ and $(h,H)$, with
masses $m_A$, $m_h$ and $m_H$ ($G_0$ turns out to be the true would-be
Goldstone boson, absorbed by the longitudinal part  of the $Z$ boson).
Here we use the  same conventions of \cite{Pilaftsis:1998dd}, to which
the reader is referred for details concerning the model [and the model
dependent factor of, for example, Eq.(\ref{thresh})].

It turns  out that the $AH$  mixing is dominated by the large CP-violating
Yukawa couplings to the top and bottom squarks, for which one has (see
again  \cite{Pilaftsis:1998dd},  where  all  the  definitions  of  the
quantities appearing in the formula below can be found)
\begin{eqnarray}
\Pi_{H_iA}(s) &=& \frac{1}{16\pi^2}\, \sum_{q=t,b}N^q_c\,\mathrm{Im}(h^q_1)
\Bigg\{\frac{r_i s_{2q}}{s_\beta v}\Delta M^2_{\tilde{q}} 
\Big[B_0(s,M^2_{\tilde{q}_1},M^2_{\tilde{q}_2})
-B_0 (0,M^2_{\tilde{q}_1},M^2_{\tilde{q}_2})\Big]
\nonumber\\
&+& \mathrm{Re}(h^q_i)\frac{s^2_{2q}}{s_\beta} 
\Big[B_0 (s,M^2_{\tilde{q}_1},M^2_{\tilde{q}_1}) + 
B_0 (s,M^2_{\tilde{q}_2},M^2_{\tilde{q}_2}) -
2B_0 (s,M^2_{\tilde{q}_1},M^2_{\tilde{q}_2})\Big] \nonumber\\ 
&+& \frac{s_{2q}}{s_\beta}\Big\{(c^2_q g^{L,q}_i + 
s^2_q g^{R,q}_i )\Big[B_0 (s,M^2_{\tilde{q}_1},M^2_{\tilde{q}_2}) -
B_0 (s,M^2_{\tilde{q}_1},M^2_{\tilde{q}_1})\Big]\nonumber\\
&+& (s^2_q g^{L,q}_i + c^2_q g^{R,q}_i ) 
\Big[B_0 (s,M^2_{\tilde{q}_2},M^2_{\tilde{q}_2}) -
B_0 (s,M^2_{\tilde{q}_1},M^2_{\tilde{q}_2})\Big]\Big\}\Bigg\}, 
\label{AHsquarks}
\end{eqnarray}
with $s=q^2$, and $B_0(s,m_1,m_2)$ the standard Passarino-Veltman function defined as
\begin{eqnarray}
B_0(s,m_1,m_2)&=& C_{\mbox{\scriptsize{UV}}}+2-\ln(m_1m_2)+\frac1s\left[(m_2^2-m_1^2)\ln\left(\frac{m_1}{m_2}\right)\right.\nonumber \\
&+&\left.\sqrt{\lambda(s,m^2_1,m^2_2)}\cosh^{-1}\left(\frac{m_1^2+m_2^2-s}{2m_1m_2}\right)\right],\nonumber \\
B_0(0,m_1,m_2)&=&C_{\mbox{\scriptsize{UV}}}+1-\ln(m_1m_2)+\frac{m_1^2+m_2^2}{m_1^2-m_2^2}\ln\left(\frac{m_2}{m_1}\right),
\end{eqnarray}
where
$C_{\mbox{\scriptsize{UV}}}=1/\epsilon-\gamma_E+\ln(4\pi\mu^2)$ is the UV cutoff 
of dimensional regularization, and $\lambda(x,y,z)=(x-y-z)^2-4yz$.

As  emphasized in  \cite{Pilaftsis:1998dd},  the tadpole  contribution
$B_0    (0,M^2_{\tilde{q}_1},M^2_{\tilde{q}_2})$   is    crucial   for
accomplishing  the  UV-finiteness   of  the  $H_iA$  self-energies  in
question.    Notice   also   that   the   $\Pi_{H_iA}(s)$   given   by
(\ref{AHsquarks})  are real,  because,  due to  the  heaviness of  the
squarks appearing in the loops,  no particle thresholds can open.  The
CP violating  self-energy transition $HA$ is then  obtained through an
orthogonal rotation of the above result, giving
\begin{equation}
\Pi_{HA}(s)=-\sin\alpha\ \Pi_{H_1A}(s)+\cos\alpha\ \Pi_{H_2A}(s).
\label{MSSMAH}
\end{equation}
Notice that $hA$ mixing is also possible, but $m_h\ll m_A,\ m_H$ 
and therefore its contribution is negligible, since one is far from 
the resonant enhancement region. 

The kinematic regime analized in this section is that where $m_A>2M_Z$
and   $\tan\beta\geq2$;   then   since   $\tan\beta\approx\tan\alpha$,
Eq.(\ref{MSSMAH}) reduces to $\Pi_{HA}(s)\approx-\Pi_{H_1A}(s)$. Also,
the rotation  angle of the scalar  top and bottom quarks  are equal to
$\pi/4$, {\it  i.e.} $s_{2q}\approx1$. As already  mentioned before we
assume          universal         squarks          soft         masses
($\tilde{M_Q}=\tilde{M_t}=\tilde{M_b}=\  M_0=.5$  TeV)  and  trilinear
couplings  ($A_t=A_b=A$ with  $|A|=1$).  Finally  $m_{H^\pm}$  will be
fixed at the value of $.4571$ TeV.

\begin{figure}[t]
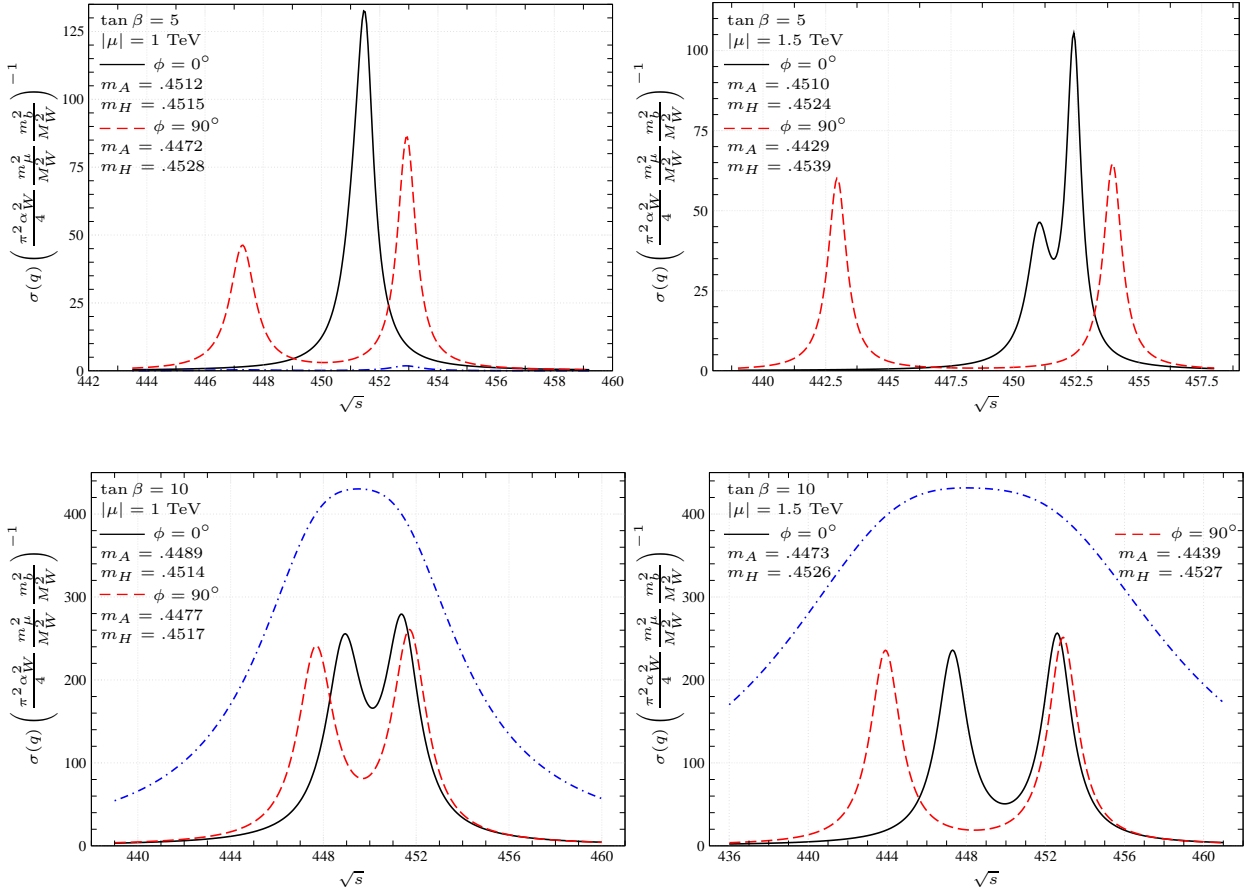

\begin{picture}(300,340)(90,-170) 
\begin{tabular}{cc}
\includegraphics[width=7.5cm]{cs-1-1-5-90.eps} \hspace{10pt} \vspace{25pt} & \includegraphics[width=7.4cm]{cs-1.5-1-5-90.eps}\\
\includegraphics[width=7.5cm]{cs-1-1-10-90.eps} \hspace{12pt} & \includegraphics[width=7.5cm]{cs-1.5-1-10-90.eps} \\ 
\end{tabular}
\put(-435,160){\tiny{$\tan\beta=5$}}
\put(-435,152.5){\tiny{$|\mu|=1$ TeV}}
\psline[linewidth=.5pt](-15.3,5.1)(-14.7,5.1)
\put(-415,143.5){\tiny{$\phi=0^\circ$}}
\put(-435,136){\tiny{$m_A=.4512$}}
\put(-435,128.5){\tiny{$m_H=.4515$}}
\psline[linewidth=.5pt,linecolor=red,linestyle=dashed](-15.3,4.25)(-14.7,4.25)
\put(-415,120){\tiny{$\phi=90^\circ$}}
\put(-435,112.5){\tiny{$m_A=.4472$}}
\put(-435,105){\tiny{$m_H=.4528$}}
\put(-470,55){\rput[lt]{90}{\tiny{$\sigma(q)\left(\frac{\pi^2\alpha^2_W}{4}\frac{m^2_\mu}{M^2_W}\frac{m^2_b}{M^2_W}\right)^{-1}$}}}
\put(-345,15){\tiny{$\sqrt s$}}
\put(-198,160){\tiny{$\tan\beta=5$}}
\put(-198,152.5){\tiny{$|\mu|=1.5$ TeV}}
\psline[linewidth=.5pt](-7.0,5.1)(-6.4,5.1)
\put(-178,143.5){\tiny{$\phi=0^\circ$}}
\put(-198,136){\tiny{$m_A=.4510$}}
\put(-198,128.5){\tiny{$m_H=.4524$}}
\psline[linewidth=.5pt,linecolor=red,linestyle=dashed](-7.0,4.25)(-6.4,4.25)
\put(-178,120){\tiny{$\phi=90^\circ$}}
\put(-198,112.5){\tiny{$m_A=.4429$}}
\put(-198,105){\tiny{$m_H=.4539$}}
\put(-233,55){\rput[lt]{90}{\tiny{$\sigma(q)\left(\frac{\pi^2\alpha^2_W}{4}\frac{m^2_\mu}{M^2_W}\frac{m^2_b}{M^2_W}\right)^{-1}$}}}
\put(-105,15){\tiny{$\sqrt s$}}
\put(-435,-17.5){\tiny{$\tan\beta=10$}}
\put(-435,-25){\tiny{$|\mu|=1$ TeV}}
\psline[linewidth=.5pt](-15.3,-1.15)(-14.7,-1.15)
\put(-415,-34){\tiny{$\phi=0^\circ$}}
\put(-435,-41.5){\tiny{$m_A=.4489$}}
\put(-435,-49){\tiny{$m_H=.4514$}}
\psline[linewidth=.5pt,linecolor=red,linestyle=dashed](-15.3,-1.95)(-14.7,-1.95)
\put(-415,-57.5){\tiny{$\phi=90^\circ$}}
\put(-435,-65){\tiny{$m_A=.4477$}}
\put(-435,-72.5){\tiny{$m_H=.4517$}}
\put(-470,-120){\rput[lt]{90}{\tiny{$\sigma(q)\left(\frac{\pi^2\alpha^2_W}{4}\frac{m^2_\mu}{M^2_W}\frac{m^2_b}{M^2_W}\right)^{-1}$}}}
\put(-345,-165){\tiny{$\sqrt s$}}
\put(-198,-17.5){\tiny{$\tan\beta=10$}}
\put(-198,-25){\tiny{$|\mu|=1.5$ TeV}}
\psline[linewidth=.5pt](-7.0,-1.15)(-6.4,-1.15)
\put(-178,-34){\tiny{$\phi=0^\circ$}}
\put(-198,-41.5){\tiny{$m_A=.4473$}}
\put(-198,-49){\tiny{$m_H=.4526$}}
\psline[linewidth=.5pt,linecolor=red,linestyle=dashed](-1.8,-1.15)(-1.2,-1.15)
\put(-30,-34){\tiny{$\phi=90^\circ$}}
\put(-50,-41.5){\tiny{$m_A=.4439$}}
\put(-50,-49){\tiny{$m_H=.4527$}}
\put(-233,-120){\rput[lt]{90}{\tiny{$\sigma(q)\left(\frac{\pi^2\alpha^2_W}{4}\frac{m^2_\mu}{M^2_W}\frac{m^2_b}{M^2_W}\right)^{-1}$}}}
\put(-105,-165){\tiny{$\sqrt s$}}
\end{picture}
\caption{\it The MSSM Higgs lineshape for different values of $|\mu|$, $\tan\beta$ and $\phi$. The black continuous curves correspond to the CP-invariant limit of the theory ($\phi=0^\circ$), while the red dashed one correspond to the case in which CP-breaking phases have been switched on (with $\phi=90^\circ$). Finally, the dashed-dotted blue curve, when present, correspond to the CP-invariant limit of the theory re-calculated for a different value of $\tan\beta$ to accommodate the mass-splitting observed in the CP-breaking case.}
\label{cross-sec90}
\end{figure}

For  relatively   big  values  of   the  CP-breaking  phase   (we  set
$\phi=90^\circ$)  our results are  shown in  Fig.\ref{cross-sec90}. 

As anticipated in the analysis carried out  in Section I, we can see that
as $\tan\beta$  decreases (upper  panels, where  $\tan\beta=5$),  
the  difference between the two line-shapes (the CP-invariant limit,
black  continuous  curves,  and  the CP-breaking  phase, red  dashed
curves) increases, the overlapping resonance get resolved and  the two mass peaks are clearly
separated. At this point one can determine directly from the lineshape
the new  pole mass  of the Higgs bosons, which turn  out to be  always in
good agreement with the  masses obtained  algebraically  by  re-diagonalizing  the  one-loop
propagator  matrix,  {\it i.e.}   by  solving  the  characteristic  equation
$\Delta(s)=0$, yielding
\begin{equation}
s_i=M^2_i-\widehat\Pi_{ii}(s_i)+\frac{\widehat\Pi_{ij}(s_i)\widehat\Pi_{ji}(s_i)}{s_i-M^2_j+\widehat\Pi_{jj}(s_i)},
\end{equation}
where  $i,j=A,H$  and  $i\ne j$ ~\cite{Carena:2001fw}.

\begin{figure}[t]
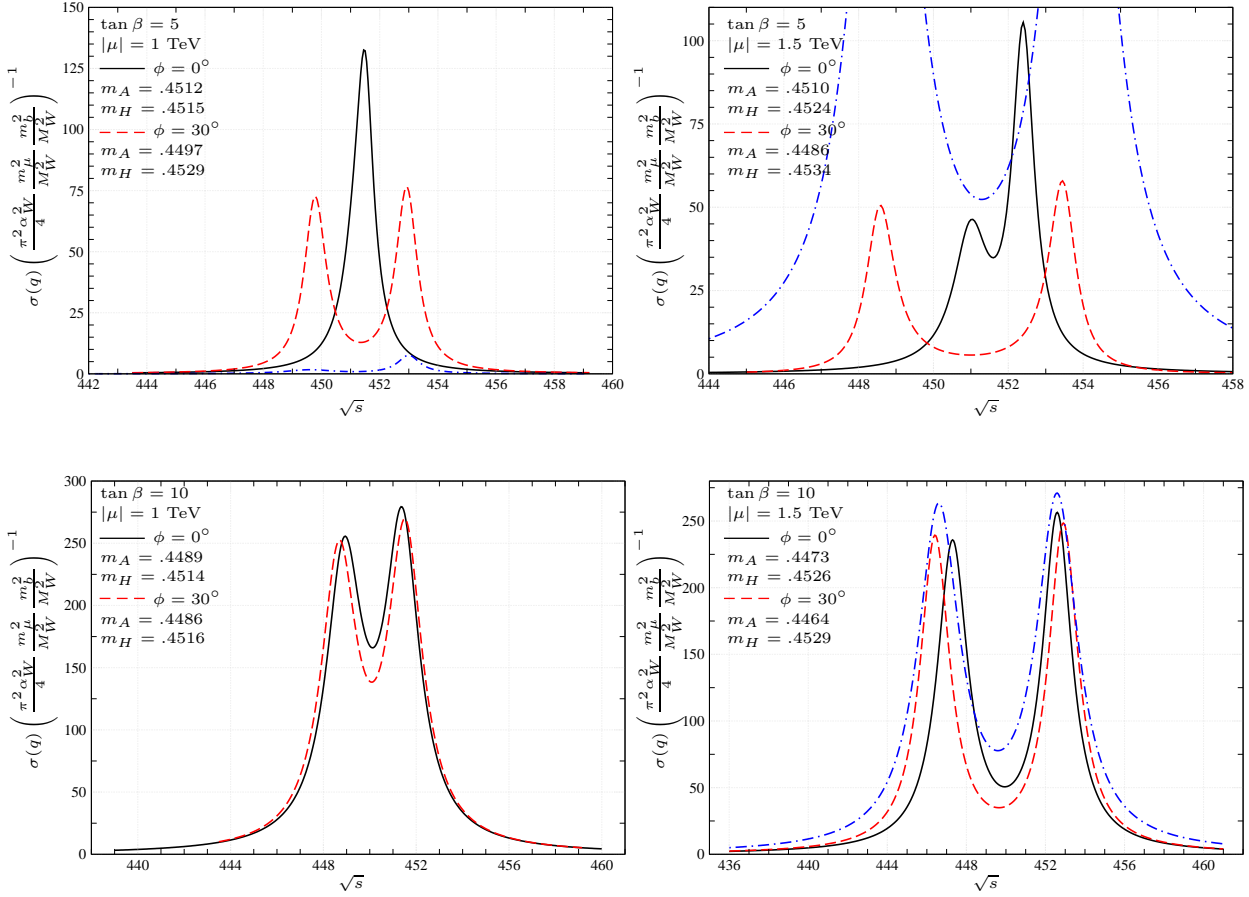

\begin{picture}(300,340)(90,-170) 
\begin{tabular}{cc}
\includegraphics[width=7.5cm]{cs-1-1-5-30.eps} \hspace{10pt} \vspace{25pt} & \includegraphics[width=7.5cm]{cs-1.5-1-5-30.eps}\\
\includegraphics[width=7.5cm]{cs-1-1-10-30.eps} \hspace{12pt} & \includegraphics[width=7.5cm]{cs-1.5-1-10-30.eps} \\ 
\end{tabular}
\put(-435,160){\tiny{$\tan\beta=5$}}
\put(-435,152.5){\tiny{$|\mu|=1$ TeV}}
\psline[linewidth=.5pt](-15.3,5.1)(-14.7,5.1)
\put(-415,143.5){\tiny{$\phi=0^\circ$}}
\put(-435,136){\tiny{$m_A=.4512$}}
\put(-435,128.5){\tiny{$m_H=.4515$}}
\psline[linewidth=.5pt,linecolor=red,linestyle=dashed](-15.3,4.25)(-14.7,4.25)
\put(-415,120){\tiny{$\phi=30^\circ$}}
\put(-435,112.5){\tiny{$m_A=.4497$}}
\put(-435,105){\tiny{$m_H=.4529$}}
\put(-470,55){\rput[lt]{90}{\tiny{$\sigma(q)\left(\frac{\pi^2\alpha^2_W}{4}\frac{m^2_\mu}{M^2_W}\frac{m^2_b}{M^2_W}\right)^{-1}$}}}
\put(-345,15){\tiny{$\sqrt s$}}
\put(-198,160){\tiny{$\tan\beta=5$}}
\put(-198,152.5){\tiny{$|\mu|=1.5$ TeV}}
\psline[linewidth=.5pt](-7.0,5.1)(-6.4,5.1)
\put(-178,143.5){\tiny{$\phi=0^\circ$}}
\put(-198,136){\tiny{$m_A=.4510$}}
\put(-198,128.5){\tiny{$m_H=.4524$}}
\psline[linewidth=.5pt,linecolor=red,linestyle=dashed](-7.0,4.25)(-6.4,4.25)
\put(-178,120){\tiny{$\phi=30^\circ$}}
\put(-198,112.5){\tiny{$m_A=.4486$}}
\put(-198,105){\tiny{$m_H=.4534$}}
\put(-233,55){\rput[lt]{90}{\tiny{$\sigma(q)\left(\frac{\pi^2\alpha^2_W}{4}\frac{m^2_\mu}{M^2_W}\frac{m^2_b}{M^2_W}\right)^{-1}$}}}
\put(-105,15){\tiny{$\sqrt s$}}
\put(-435,-17.5){\tiny{$\tan\beta=10$}}
\put(-435,-25){\tiny{$|\mu|=1$ TeV}}
\psline[linewidth=.5pt](-15.3,-1.15)(-14.7,-1.15)
\put(-415,-34){\tiny{$\phi=0^\circ$}}
\put(-435,-41.5){\tiny{$m_A=.4489$}}
\put(-435,-49){\tiny{$m_H=.4514$}}
\psline[linewidth=.5pt,linecolor=red,linestyle=dashed](-15.3,-1.95)(-14.7,-1.95)
\put(-415,-57.5){\tiny{$\phi=30^\circ$}}
\put(-435,-65){\tiny{$m_A=.4486$}}
\put(-435,-72.5){\tiny{$m_H=.4516$}}
\put(-470,-120){\rput[lt]{90}{\tiny{$\sigma(q)\left(\frac{\pi^2\alpha^2_W}{4}\frac{m^2_\mu}{M^2_W}\frac{m^2_b}{M^2_W}\right)^{-1}$}}}
\put(-345,-165){\tiny{$\sqrt s$}}
\put(-198,-17.5){\tiny{$\tan\beta=10$}}
\put(-198,-25){\tiny{$|\mu|=1.5$ TeV}}
\psline[linewidth=.5pt](-7.0,-1.15)(-6.4,-1.15)
\put(-178,-34){\tiny{$\phi=0^\circ$}}
\put(-198,-41.5){\tiny{$m_A=.4473$}}
\put(-198,-49){\tiny{$m_H=.4526$}}
\psline[linewidth=.5pt,linecolor=red,linestyle=dashed](-7.0,-1.95)(-6.4,-1.95)
\put(-178,-57.5){\tiny{$\phi=30^\circ$}}
\put(-198,-65){\tiny{$m_A=.4464$}}
\put(-198,-72.5){\tiny{$m_H=.4529$}}
\put(-233,-120){\rput[lt]{90}{\tiny{$\sigma(q)\left(\frac{\pi^2\alpha^2_W}{4}\frac{m^2_\mu}{M^2_W}\frac{m^2_b}{M^2_W}\right)^{-1}$}}}
\put(-105,-165){\tiny{$\sqrt s$}}
\end{picture}
\caption{\it Same as before but for the CP-breaking phase which is now set to $\phi=30^\circ$. Notice that in the upper left panel, the same masses could have been also reached from a CP-invariant theory with $\tan\beta\sim25$, which however would give a much higher cross section which is not shown.}
\label{cross-sec30}
\end{figure}

In particular, notice that when $\vert\mu\vert=1.5$ TeV(upper-right panel)
the mass  splitting $\delta  m\sim11$ GeV and  cannot be  explained in
terms of  CP-invariant radiative corrections  only, no matter  how big
$\tan\beta$ is chosen. In the  remaining panels, $\delta m$ allows for
smaller  (upper-left  panel,   $\tan\beta\sim2.5$)  or  larger  (lower
panels, left $\tan\beta\sim20$  and right $\tan\beta\sim30$) values of
$\tan\beta$ which  are consequently  plotted in th  blue dashed-dotted
curves. Also in  these cases one can distinguish the  CP-invariant and
CP-breaking  phase of  the theory  due  to the  very different  energy
behavior of  the cross section; also  notice that too  lower values of
$\tan\beta$ predicts,  in the parameter  range chosen here, a  too low
mass  for the  lightest  Higgs, and  therefore  would be  in any  case
excluded  ({\it e.g.} when  $\tan\beta\sim2.5$, $m_h=94.8$  GeV).  For
smaller values of the CP-breaking  phase $\phi$ we get a
somewhat different  situation. In Fig.\ref{cross-sec30},  the same set
of plots  as before is shown  for $\phi=30^\circ$: for  small values of
$\tan\beta$ (upper  panels) it is  still possible to  discriminate the
phase  of the  theory,  while  for larger  values  (lower panels)  the
effects  of  $A-H$ mixing  vanishes  much  more  rapidly than  in  the
previous case.

\section{Conclusions}

The CP-even and CP-odd neutral Higgs scalars, $H$ and $A$, respectively,
together with their charged companions and the ``standard'' Higgs boson, 
constitute the extended scalar sector of generic  
two-Higgs doublet models, and appear naturally in supersymmetric
extensions of the SM. Therefore it is expected that their discovery
and subsequenet study of their fundamental physical properties 
should be of central importance in the next decades.
In addition to the LHC, where their primary discovery might take place,
further detailed studies of their characteristics
have been proposed in recent years, most notably  
in the context of muon-\cite{Cline:1994tk,Marciano:1998ue} 
and photon colliders~\cite{Ginzburg:2001ph,Choi:2004kq}.

Of particular interest 
is the possibility of encountering CP violation in the  Higgs sector
of two-Higgs doublet models, induced by the one-loop mixing of $H$ and $A$.
The latter is produced 
due to CP-violating (tree-level) interactions of each one of these two Higgs bosons
with particles such as scalar quarks or Majorana neutrinos, to name a few,
which will circulate in the (otherwise vanishing) $H$-$A$ transition amplitude
(Fig.1).

A most  appealing feature of  the $H$-$A$ system  is that, due  to their
relative  small mass-difference,  possible  CP-mixing effects  undergo
resonant  enhancement~\cite{Pilaftsis:1997dr}.   Such  effect  may  be
observed  in  appropriately  constructed  left-right  asymmetries,  as
proposed in  \cite{Pilaftsis:1997dr}. Since these  observables are odd
under  CP, a  non-vanishing  experimental result  would constitute  an
unequivocal signal of CP violation.

In  this paper  we  have  explored the  possibility  of detecting  the
presence of CP-mixing  between $H$ and $A$ through  the detailed study
of  the  cross-section   of  the  $s$-channel  process  $\mu^+\mu^-\to
A^*,H^*\to f\bar f$  as function of the center-of-mass  energy.  
Although the lineshape is a CP-even quantity, there are certain 
characteristics that signal the presence of a 
CP-violating mixing. 
In the context of the photon colliders, Ref.\cite{Choi:2004kq} also 
discusses both CP-even and CP-odd observables in terms of photon
polarizations.

As    has     been    explained    in    a     series    of    papers~
\cite{Barger:1996jm,Binoth:1997ev,Krawczyk:1997be,Grzadkowski:2000fg,Blochinger:2002hj},
the  high-resolution scanning  of  the lineshape  of  this process  is
expected   to  reveal   two   relatively  closely   spaced,  or   even
superimposed, resonances,  corresponding to the  nearly degenerate $H$
and $A$.   As was recognized in \cite{Carena:2001fw},  the presence of
CP-mixing between $H$ and $A$  modifies the position of the pole mass,
and tends to  push the two resonances further  apart.  Of course, this
fact  by itself  could  not serve  as  a signal  for  the presence  of
CP-mixing;  one   could  simply  envisage  the   situation  where  all
fundamental interactions are CP-conserving  and there is no CP-mixing,
and the masses  of $H$ and $A$ assume simply  from the beginning their
shifted values.

The result  of our  analysis is very  positive and represents,  in our
opinion, a significant step in  the search of the CP-properties of two
scalar-doublet models: {\it  either the mass-splitting of the  H and A
Higgs boson, in a CP-mixing  scenario, cannot be accounted for 
in  absence  of  CP-mixing, and/or  the  detailed  energy
dependence  of  the line-shape  allows  to  discriminate between  both
scenarios}.  

Even though  in our analysis we  have considered only the  MSSM due to
its phenomenological  relevance, we  however expect that  our analysis
will hold in general for  two Higgs doublet models, independently from
the specific type of mechanism which will give rise to the CP-breaking
loops  of  Fig.\ref{feyndiag} (a).  For  example  similar results  are
expected  in the  case  of a  two  Higgs doublet  model  in which  the
particles  circulating in  the  loop are  three  generations of  heavy
Majorana neutrinos  $N_i$ ($i=1,2,3$)  for which the  mixing amplitude
will be given by \cite{Pilaftsis:1997dr}
\begin{eqnarray}
& &\Pi_{AH}(s)=\ -\frac{\alpha_w \,s}{4\pi}\chi_{Au}\chi_{Hu}\sum\limits_{j>i}^{3} \mathrm{Im}(C^2_{ij})
\sqrt{\lambda_i\lambda_j}\left[ B_0(s/M^2_W,\lambda_i,\lambda_j)+\, 2B_1(s/M^2_W,\lambda_i,\lambda_j)\right],\nonumber \\
& & B_1(s,m_i,m_j) = \frac{m_j^2 - m_i^2}{2s}B_0(s, m_i, m_j) - B_0(0,m_i, m_j) - \frac12B_0(s, m_i, m_j),
\end{eqnarray}
with $\lambda_i = m^2_i/M^2_W$, and $\mathrm{Im}(C^2_{ij})$ playing the role of the MSSM CP-breaking phase $\phi$.

Thus,  the combined study of CP-odd  observables, such as
left-right asymmetries, together with the $H$-$A$ lineshape, may furnish
a  powerful  panoply for  attacking  the  isssue  of CP  violation  in
two-Higss doublet models.

\medskip
\medskip

{\it Acknowledgments:} This research has been supported 
by the Spanish MEC and European FEDER, under the Grant FPA 2005-01678.
 Useful correspondence with A. Pilaftsis and J. S. Lee is gratefully acknowledged.
We thank Peter Zerwas for bringing reference \cite{Choi:2004kq} to our
attention. JaxoDraw \cite{Binosi:2003yf} has been used.

\end{document}